\documentclass[aps,pre,twocolumn,showpacs,floatfix,preprintnumbers,amsmath,amssymb]{revtex4-1}
\usepackage{amsmath}
\usepackage{amssymb}
\usepackage{graphicx}
\usepackage{epsfig}
\usepackage{dcolumn}
\usepackage{bm}
\usepackage{array}
\usepackage{xcolor}

\begin{document}
\title{Scaling in the eigenvalue fluctuations of correlation matrices}
\author{Udaysinh T. Bhosale}
\email{udaybhosale0786@gmail.com}
\author{S. Harshini Tekur}
\email{harshini.t@gmail.com}
\author{M. S. Santhanam}
\email{santh@iiserpune.ac.in}
\affiliation{Indian Institute of Science Education and Research, Dr. Homi Bhabha Road, Pune 411 008, India}
\begin{abstract}
The spectra of empirical correlation matrices, constructed from multivariate data,
are widely used in many areas of sciences, engineering and social sciences as a tool
to understand the information contained in typically large datasets. In the last
two decades, random matrix theory-based tools such as the nearest neighbour eigenvalue spacing
and eigenvector distributions have been employed to extract the significant
modes of variability present in such empirical correlations. In this work, we present
an alternative analysis in terms of the recently introduced spacing ratios, which 
does not require the cumbersome unfolding process. It is shown that the higher order
spacing ratio distributions for the Wishart ensemble of random matrices, characterized
by the Dyson index $\beta$, is related to the first order spacing ratio distribution with a modified value
of co-dimension $\beta'$. This scaling is demonstrated for Wishart ensemble and also
for the spectra of empirical correlation matrices drawn from the observed stock market 
and atmospheric pressure data. Using a combination of analytical and numerics, such
scalings in spacing distributions are also discussed.
\end{abstract}

\maketitle

\section{Introduction} 

Large multivariate datasets are commonly encountered in many areas of
sciences \cite{Wishart1928,Kwapien2012}, engineering \cite{RMTApplications} 
and social sciences \cite{BartholomewBook}. Some common examples include 
the data generated from the financial markets \cite{rmt-fin}, atmospheric and climate 
parameters \cite{Santhanam2001} and communication networks \cite{Barlowe2008}. 
Analysis of the spectra of empirical correlation matrices constructed 
from large data sets provides detailed and
graded information about the systems being studied. In the last two decades,
tools and results from random matrix theory (RMT) have been widely applied to make sense of
the information provided by detailed spectra, namely, the eigenvalues and
the eigenvectors, of the empirical correlation matrices \cite{wilksbook,jbun}.
Originally, RMT was conceived as a model for the energy spectra of complex many-body 
quantum systems such as nuclei and atoms \cite{rmt1,GuhrReport,LivanBook}. These novel applications of RMT
have expanded its scope well outside of its original domain of quantum spectra.

The eigenvalues $E_i, i=1,2,\ldots,N$ of the empirical correlation matrix of
order $N$ are positive definite, i.e., $E_i \ge 0$.
Typically, the corresponding eigenmodes fall in two broad groups;
({\it i}) eigenmodes of the top and bottom few eigenvalues (in magnitude) that carry 
most of the information embedded in the original dataset
({\it ii}) the bulk of rest represents random correlations.
It is the latter group that displays a broad agreement with random matrix based results.
For instance, the bulk of the correlation matrix spectra obtained from the 
time series of the largest stocks in the United States, including the ones that
make up the S\&P index, was shown to be in agreement
with the random matrix averages \cite{ecm-rmt,Sitabhra2007,Plerou2002,Laloux1999}, and some studies have argued
that they contain finer correlated structures \cite{livan1,Kwapien2005}.
The density of the bulk of eigenvalues follows Marcenko-Pastur distribution \cite{Marcenko67}
and can be used to identify the top eigenvalues that carry significant information.
The analysis of eigenvectors, in terms of its agreement with Porter-Thomas distribution \cite{ptdist}, 
indicates the stocks that are strongly correlated \cite{Laloux1999}. A similar approach for 
the analysis of atmospheric data can distinguish physically relevant modes of atmospheric variability 
from the those that are noisy \cite{Santhanam2001}. 
By now, many applications \cite{akemann2011oxford,Novembre2008,Patterson2006,Akemann1997,Johansson2000} 
ranging from biology \cite{rmt-bio}, image processing \cite{fukunaga1990} and network 
traffic \cite{rmt-engg} abound.

An often demonstrated property of the bulk of eigenvalues is that the
distribution of the spacings $s_i=E_{i+1}-E_i, i=1,2,\dots$, between consecutive
eigenvalues (after unfolding) follows the celebrated
Wigner distribution, $P(s) = (\pi/2) s \exp(-\pi s^2/4)$ \cite{rmt1}. This signifies level repulsion, the
tendency of the eigenvalues to repel one another.
This continues to be a popular test for RMT-like behaviour,
especially for the claim that spectral fluctuations of empirical correlation
matrices display universal characteristics irrespective of the dataset or system
considered for analysis. A major impediment to computing the spacing distribution
is the requirement to unfold the eigenvalues, a somewhat unreliable numerical procedure that
approximately separates the system dependent eigenvalue properties from the generic ones.
This problem can be circumvented by considering the ratio of consecutive spacings,
$r_i = s_{i+1}/s_i, i=1,2,\dots$,
which are independent of local eigenvalue density and hence does not depend on
the system \cite{huse,atas1,atas2}. In this work, new scaling properties relating to the 
distribution of higher order spacings and spacing ratios to the nearest neighbour 
spacing properties are demonstrated.

The elements of the empirical correlation matrix represent the pair-wise Pearson
correlation among the $N$ variables, each one being a time series of
length $T$. From the point of view of random matrix theory, correlation matrices
fall within the class of Laguerre-Wishart ensemble \cite{lwe} of random matrix
theory represented by $W=D_R D_R^S$, where $D_R$ represents the standardized data matrix
of order $N$ by $T$ with real, complex or quaternion elements depending on the Dyson index $\beta=1,2,4$
of the ensemble and $X^S$ represents self-dual operation on matrix $X$.
For the Laguerre-Wishart ensemble indexed by $\beta$ the random matrix
average for the spacing ratios is not yet known, though in the limit of matrix
size $N \to \infty$, it is well-approximated by that for the Gaussian ensembles given by
\begin{equation}
P(r) = C_{\beta} ~ \frac{(r + r^2)^{\beta}}{(1+r+r^2)^{1+\frac{3}{2}\beta}},
\label{ratio_goe}
\end{equation}
where $C_{\beta}$ is a constant whose form is given in Ref.\cite{atas2}.
However, results for spacing ratios and spacing distributions beyond the nearest
neighbours are not yet known. The higher order spacing statistics provide a finer
test for the universality of spectral fluctuations. Secondly, the deviations from 
these will quantify the time-scales up to which random matrix type universality
can be expected to be valid in empirical cases. Indeed, long-range correlations 
such as $\Delta_3$ statistics have indicated limitations of RMT assumptions at longer
time scales \cite{Santhanam2001,Plerou2002,rmt-engg}.


The structure of the paper is as follows: In Sec.~\ref{sec:higherratios} a scaling relation is given for the 
higher order spacing
ratios for the Wishart ensemble of random matrices. In Sec.~\ref{sec:results}, this scaling relation has been tested on various
systems which include the spectra of the Wishart random matrix ensembles and the spectra of empirical correlation matrices drawn
from the stock market and atmospheric data set. This relation has also been shown to hold good analytically for the first few orders for the
spacing distribution. Finally, in Sec.~\ref{sec:conclusions} this work is summarized and concluded.

\begin{figure}[t]
\includegraphics*[width=3.4in]{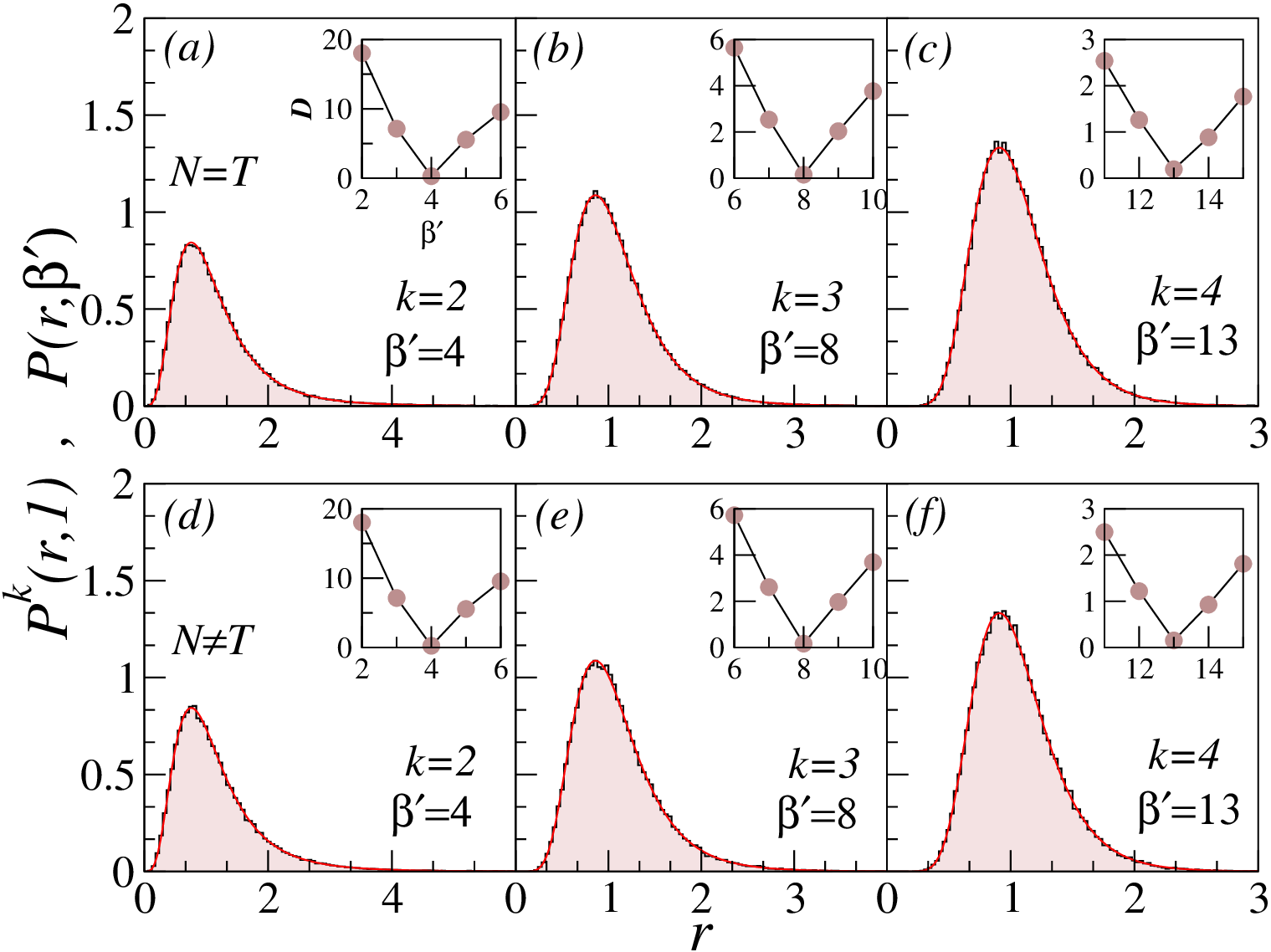}
\caption{The histograms are the $k$-th spacing ratio distribution for the spectra of 
random Wishart matrix for $\beta=1$ with $N=T=40000$ (a-c), and $N=20000,
T=30000$ (d-f). The computed histograms display a good agreement with $P(r,\beta')$ shown 
as solid line. In this, $\beta'$ is given by Eq. \ref{eq3}. Inset shows that the minima 
in $D(\beta')$ corresponds to the value of $\beta'$ predicted by Eq. \ref{eq3}.}
\label{wishartb1}
\end{figure}

\begin{figure}[t]
\includegraphics*[width=3.4in]{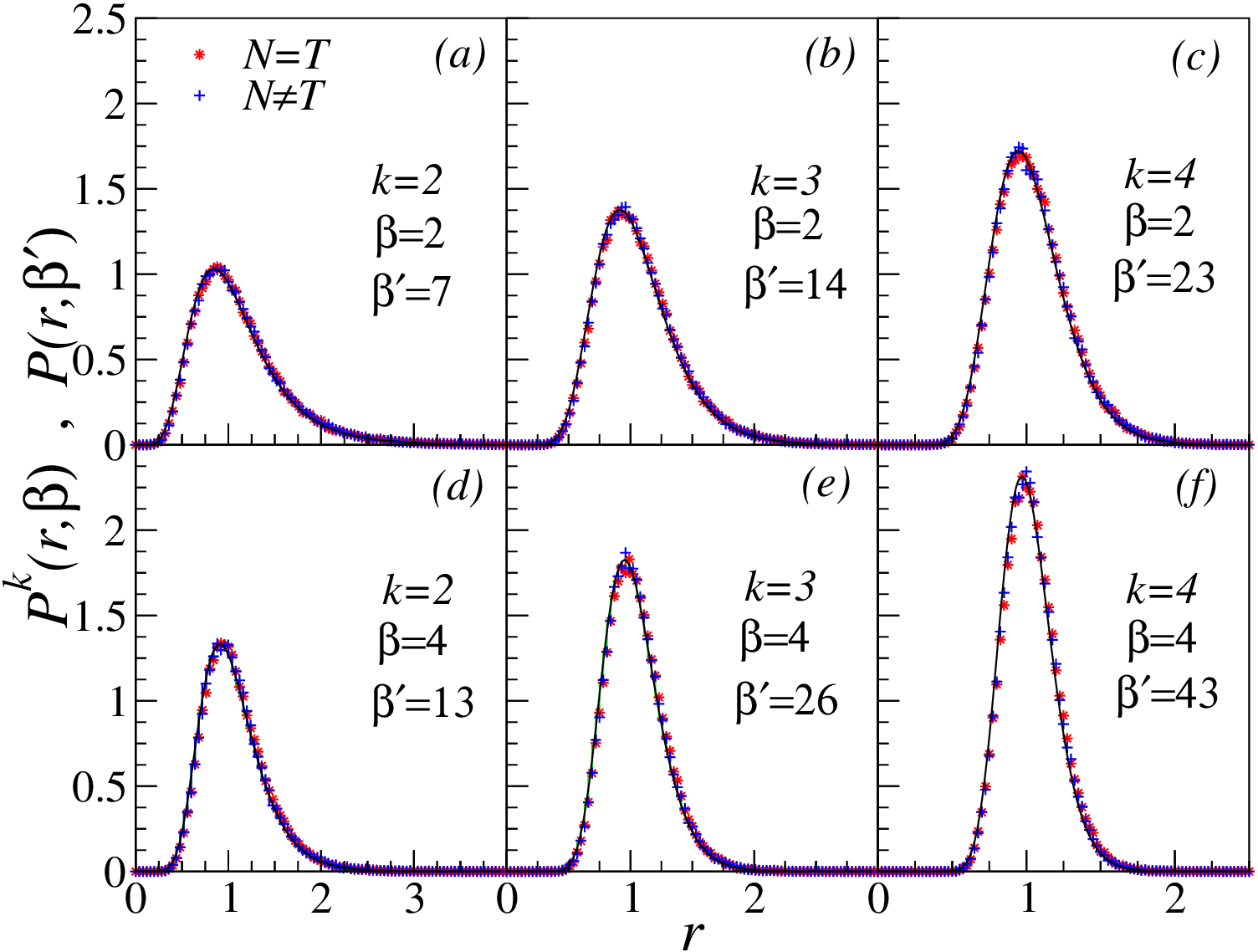}
\caption{The histograms are the $k$-th spacing ratio distribution for the spectra of
random Wishart matrix with (a-c) $\beta=2$ and (d-f) $\beta=4$.
For the $N=T$ case, $N=T=20000$; and for $N\ne T$ case, $N=10000$ and $T=20000$.
The computed histograms display a good agreement with $P(r,\beta')$ shown
as solid line ($\beta'$ given by Eq. \ref{eq3}).}
\label{wishartb24}
\end{figure}

\section{Distribution of higher order spacing ratios}
\label{sec:higherratios}

Consider a sample correlation matrix $C$ of order $N >>1$, constructed from time series
of $N$ variables and each of length $T$, whose ordered eigenvalues 
are $E_1 \le E_2 \le \dots \le E_N$. The density of eigenvalues are
given, in the limit $T \ge N >> 1$, by the Marcenko-Pastur distribution, which predicts
an upper and a lower bound for the eigenvalues \cite{Marcenko67}.
In nearly all of the earlier studies involving spectra of empirical correlation
matrices, eigenvalue density and spacing distributions had
been widely studied \cite{Santhanam2015}.
In contrast, in this work, we study the $k$-th order spacing ratios defined by,
\begin{align}
r_i^{(k)} & = \frac{s_{i+k}^{(k)}}{s_{i}^{(k)}} = \frac{E_{i+2k}-E_{i+k}}{E_{i+k}-E_i},
\label{hosr}
\end{align}
where $k$-th order spacings can also be defined as $s^{(k)} = s_{i+k}-s_i$.
If $k=1$, this reduces to the standard nearest neighbour spacing ratio.
The distribution of $k$-th order spacing ratio is denoted by
$P^k(r,\beta)$, where $\beta=1,2,4$ is the co-dimension that represents the Wishart ensemble.
Note that the higher order spacing ratios are not uniquely defined \cite{atas2}. In Eq. \ref{hosr} we take
them such that no common spacings are shared between the numerator and denominator.
Using Eq. \ref{hosr}, the main result of this paper can be stated as
follows: for the random matrix of order $N>>1$ and $T \ge N >> 1$, from Wishart ensemble with
$\beta=1, 2$ and 4, the $k$-th order spacing ratio distribution are related to the
nearest neighbour $(k=1)$ spacing ratio distribution statistics by
\begin{align}
P^k(r,\beta) & = P(r,\beta'), ~~~~~\beta=1,2,4, \label{eq2}\\
\beta' & =  \frac{k(k+1)}{2}\beta+(k-1),  ~~~~k \geq 1.
\label{eq3}
\end{align}
In this, $4 \le \beta'< \infty$, though for these values of $\beta'$ explicit
matrix forms for the Wishart ensemble are not known. Though the eigenvalue 
density given by Marcenko-Pastur distribution
depends on the ratio $T/N$ \cite{Marcenko67}, the statistics of fluctuations represented by Eq. \ref{eq3} can be
expected to be independent of $T/N$. Similar scaling relation had been 
postulated for the spacing distribution of 
Gaussian ensembles \cite{porter_abulmagd} and its generalizations \cite{Forrester2009},
and recently numerical 
evidence was provided for the ratios \cite{Harshini2018a}.
Both $P^k(r,\beta)$ and $P(r,\beta')$ have identical functional forms and the modified
parameter $\beta'$ depends on the order $k$ of the spacing ratio and co-dimension $\beta$.
Further, we present strong numerical evidence from Wishart matrices as well as
from empirical correlation matrix spectra computed from observed 
the stock market and atmospheric data.

A rigorous proof of Eqs.~(\ref{eq2}-\ref{eq3}) is mathematically challenging but we give an intuitive argument 
why $\beta'$ should be greater than $\beta$. The eigenvalues of Wishart ensemble has correspondence with the 
charged particles of a two-dimensional coloumb gas \cite{Forresterbook}. In this physical picture, the degree of 
repulsion between eigenvalue pairs beyond the nearest neighbours is greater than that for consecutive eigenvalue 
pairs. Hence, it appears physically reasonable to expect that $\beta'$ for the case of $k>1$ is greater than 
$\beta$ for $k=1$. For the special case of $k=2$ in the context of circular orthogonal ensemble ($\beta=1$) of 
RMT, a limited analytical proof was derived in Ref. \cite{MehtaDyson}. Thus, Eqs. ~(\ref{eq2}-\ref{eq3})
represents a generalization of this result for the spacing ratios of the Wishart ensemble. In the next section 
we apply the scaling relation in Eqs.~(\ref{eq2}-\ref{eq3}) to the spectra of various systems.

\begin{table}[]
\begin{tabular}{|c|c|c|c|c|c|c|c|}
\toprule

$~~\beta ~~$ & $~~k~~$ & $~~\beta' ~~$  & $\langle r\rangle_{th}$   &  $\langle r\rangle_{w}$ & $\langle r\rangle_{w}$ & $\langle r\rangle_{atm}$  & $\langle r\rangle_{fin}$  \\
 & &  &  &  $(n=m)$ & $(n\neq m)$ &  & \\ \hline
  & 2 & 4  & 1.174 & 1.177 & 1.176        & 1.214 & 1.330 \\[3pt]
1 & 3 & 8  & 1.085 & 1.086 & 1.085        & 1.123 & 1.237 \\[3pt]
  & 4 & 13 & 1.052 & 1.053 & 1.052        & 1.089 & 1.211 \\
\hline
\end{tabular}
\caption{Mean values of spacing ratio for Wishart matrix $\langle r\rangle_{w}$, 
atmospheric correlations data $\langle r\rangle_{atm}$ and stock market 
correlations $\langle r\rangle_{fin}$. The expected theoretical value is
$\langle r\rangle_{th}$.}
\label{table1}
\end{table}


\section{Results}
\label{sec:results}
\subsection{Random matrix spectra}

Now, we consider the spectra obtained from an ensemble of Wishart matrices with $\beta=1$
and test the validity of the Eq. \ref{eq2}-\ref{eq3} by computing the higher order spacing ratios.
In Fig. \ref{wishartb1}, the $k$-th order spacing ratio distributions are shown
as histograms for two cases, namely, $N=T$ and $N \ne T$. The validity of the scaling
in Eq. \ref{eq2} can be clearly inferred from the excellent agreement of the
histogram with a solid curve representing $P(r,\beta')$, where $\beta'$ given by Eq. \ref{eq3}.
An additional layer of quantitative verification can be performed as follows.
Let the cumulative distributions corresponding to the computed histogram $P^k(r_i,\beta)$
and $P(r, \beta')$ be represented, respectively, by $I^k(r_i,\beta)$ and $I(r_i,\beta')$.
In Eq. \ref{eq3}, $\beta'$ is treated as a tunable parameter and the difference
between cumulative distributions 
\begin{align}
D(\beta')=\sum_i \left|{I}^k(r_i,\beta) - I(r_i,\beta') \right|,
\end{align}
is computed. The minima of $D(\beta')$ is the best value of $\beta'$ that fits
the histogram data. As seen in the insets in Fig. \ref{wishartb1}, the minima
in $D(\beta')$ precisely coincides with the value of $\beta'$ postulated in Eq. \ref{eq3}.

The results displayed in Fig. \ref{wishartb24} show that the higher order spacing ratio
distributions computed from the spectra from Wishart matrices with $\beta=2$ and 4
are consistent with the scaling relation postulated in Eq. \ref{eq2}-\ref{eq3}.
The elements of Wishart matrices with $\beta=2$ and 4 are, respectively, complex
numbers and quaternions and empirical correlations with such elements are
rarely encountered in practice.
The symbols in this figure represent the histograms and solid curves represent
$P(r,\beta')$. The results are shown for both $N=T$ and $N \ne T$ and, as anticipated,
the agreement with Eqs. \ref{eq2}-\ref{eq3} is good irrespective of the relative
values of $N$ and $T$. Another form of evidence in Table \ref{table1} for the
mean ratio $\langle r \rangle$ shows a good agreement between the theoretically
expected value based on Eqs. \ref{eq2}-\ref{eq3} and that obtained from computed
Wishart spectra.

\subsection{Stock market and atmospheric data}

Next, we demonstrate the validity of the scaling relation Eq. \ref{eq2}-\ref{eq3} for
the spectra of empirical correlation matrices drawn from two different domains, namely,
the stock market and atmospheric data set. To begin with, the data of the time series
of stocks that are part of the S\&P500 index for the years 1996-2009 is considered \cite{livan1}.
This dataset continues to be extensively used to understand the ramifications
of how an RMT-based approach might work in the context of empirical correlation matrices.
The data consists of daily (log) returns for $T=3400$ days for $N=396$ assets.
The elements of the correlation matrix denote the Pearson correlation between pairs
of stocks averaged over time. Note that $T \ge N$ implying that the correlations can
be assumed to have converged. The statistical properties of its spectra have
been reported in \cite{Laloux1999,ecm-rmt,Plerou2002,Sitabhra2007}. 

\begin{figure}[t]
\includegraphics*[width=3.4in]{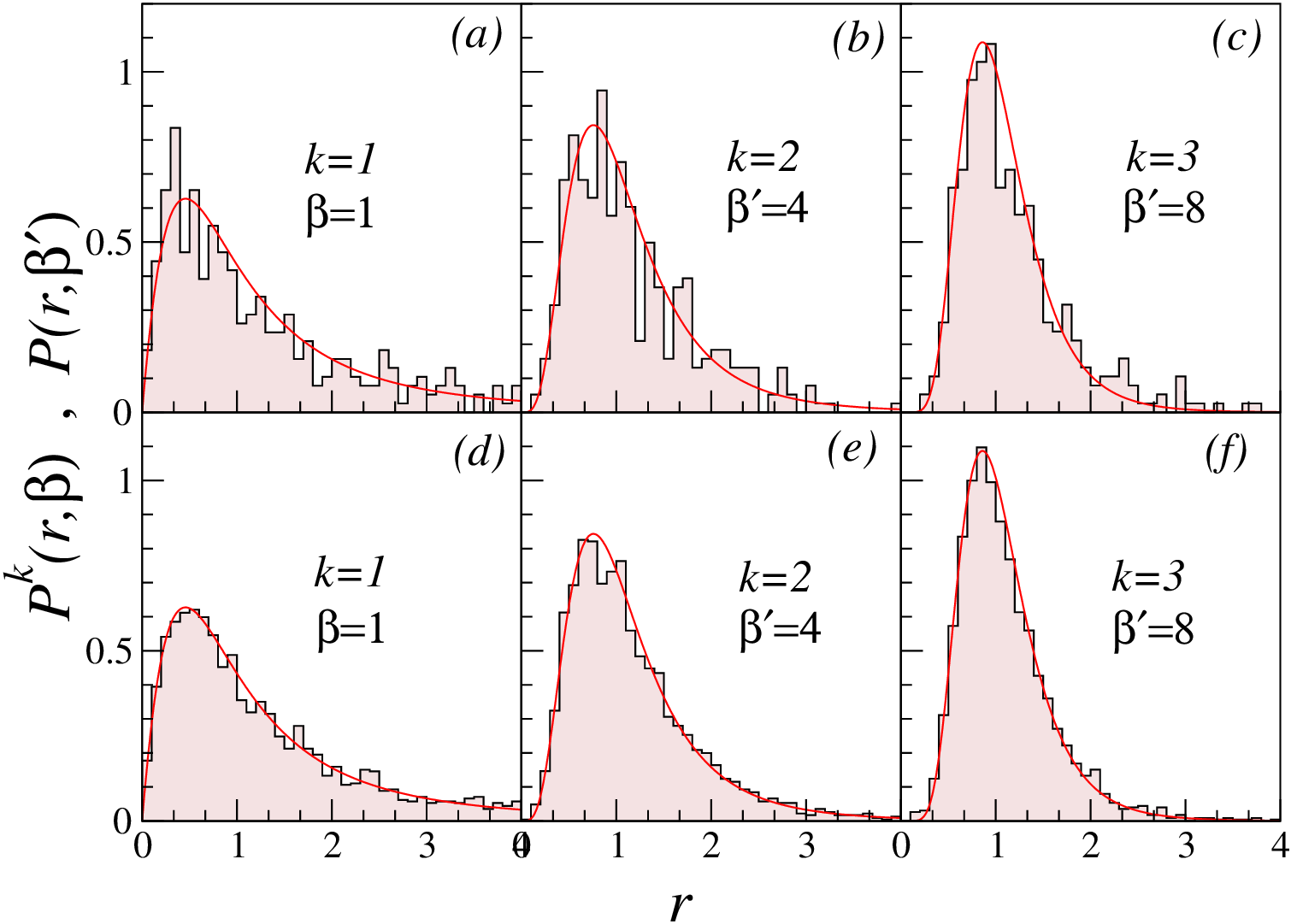}
\caption{The histograms are the $k$-th order spacing ratio distribution for
the spectra of correlation matrix (a-c) from S\&P500 stock market data and (d-f)
from mean sea level pressure data. The computed histograms display a
good agreement with $P(r,\beta')$ shown as solid line. In this, $\beta'$ is given by Eq. \ref{eq3}.}
\label{atm_fin}
\end{figure}

In Fig. \ref{atm_fin}(a-c), we display the spacing ratio distribution for various
orders. Fig. \ref{atm_fin}(a) shows the nearest neighbour spacing ratio distribution
and it agrees with the analytical result in Eq. \ref{ratio_goe} obtained for
the case of Gaussian Orthogonal Ensemble \cite{atas1}. The higher order spacing ratio distributions
are displayed in Fig. \ref{atm_fin}(b,c) and we notice a good agreement with the
postulated theoretical distribution $P(r,\beta')$, with $\beta'$ as given by 
Eq. \ref{eq3}.

\begin{figure}[b]
\includegraphics*[width=3.4in]{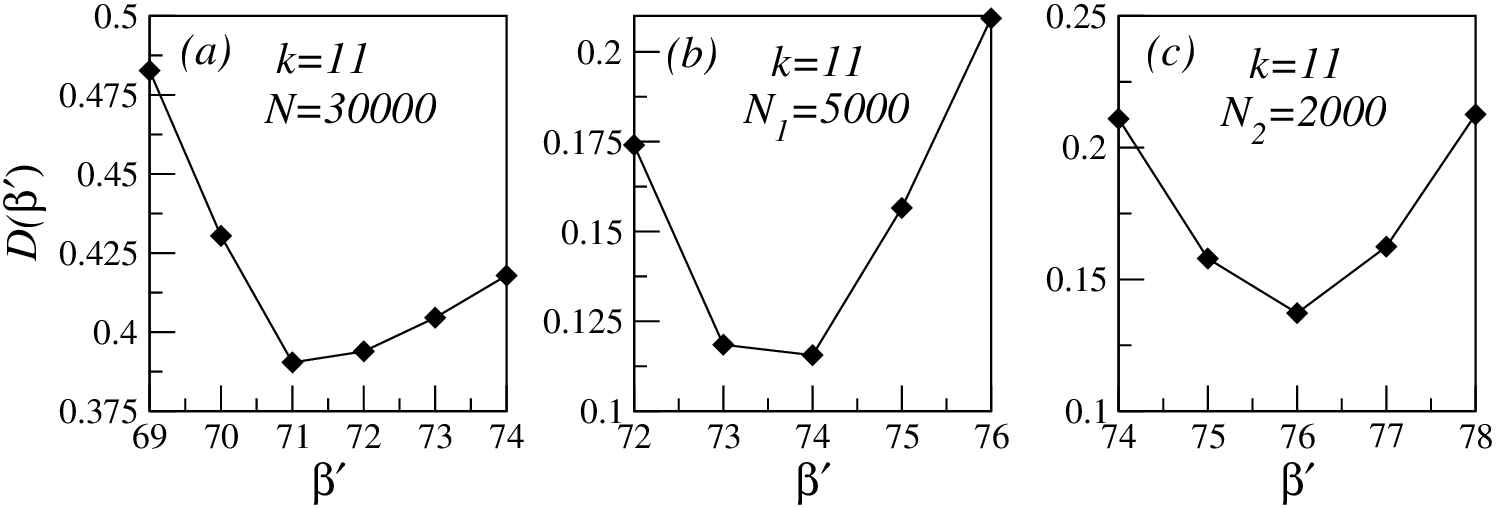}
\includegraphics*[width=3.435in]{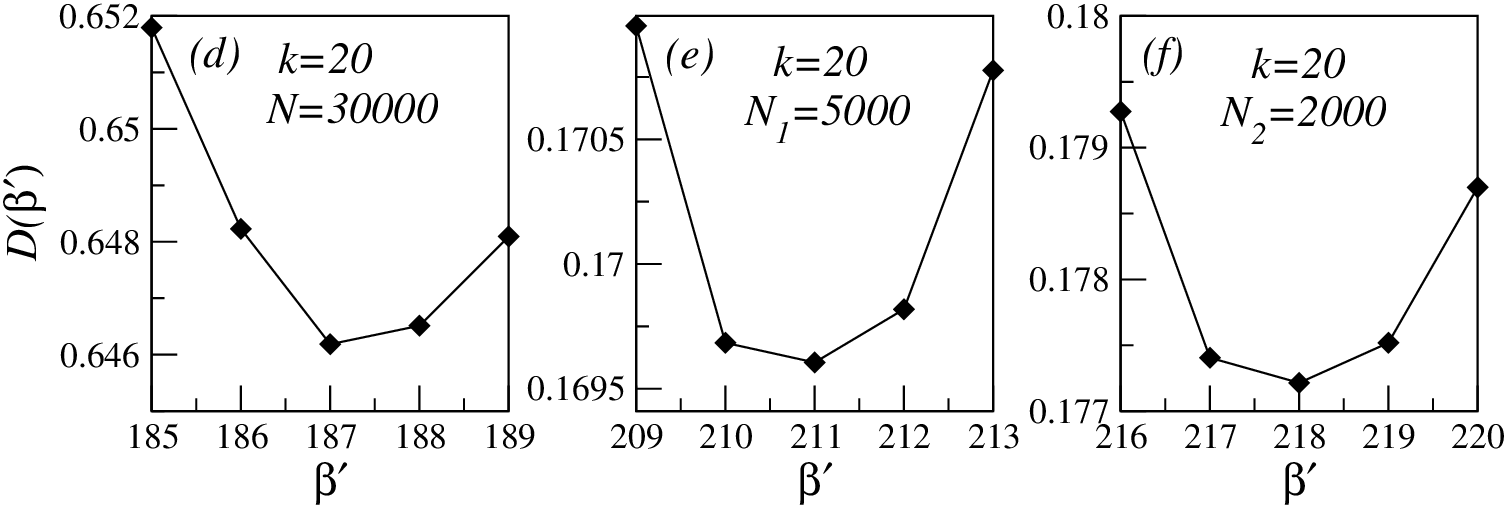}
\caption{$D(\beta')$ is plotted for $k=11$ (top panel) and $k=20$ (bottom panel) and for (a,d) $N=30000$, 
(b,e) $N_1=5000$, (c,f) $N_2=2000$ eigenvalues, for matrices of dimensions $N=T=30000$. $N_1$ and $N_2$ 
represent a subset of eigenvalues taken from the bulk of the spectrum. 
%
}
\label{fig:k11convergence}
\end{figure}

\begin{figure}[t]
\includegraphics*[width=3.4in]{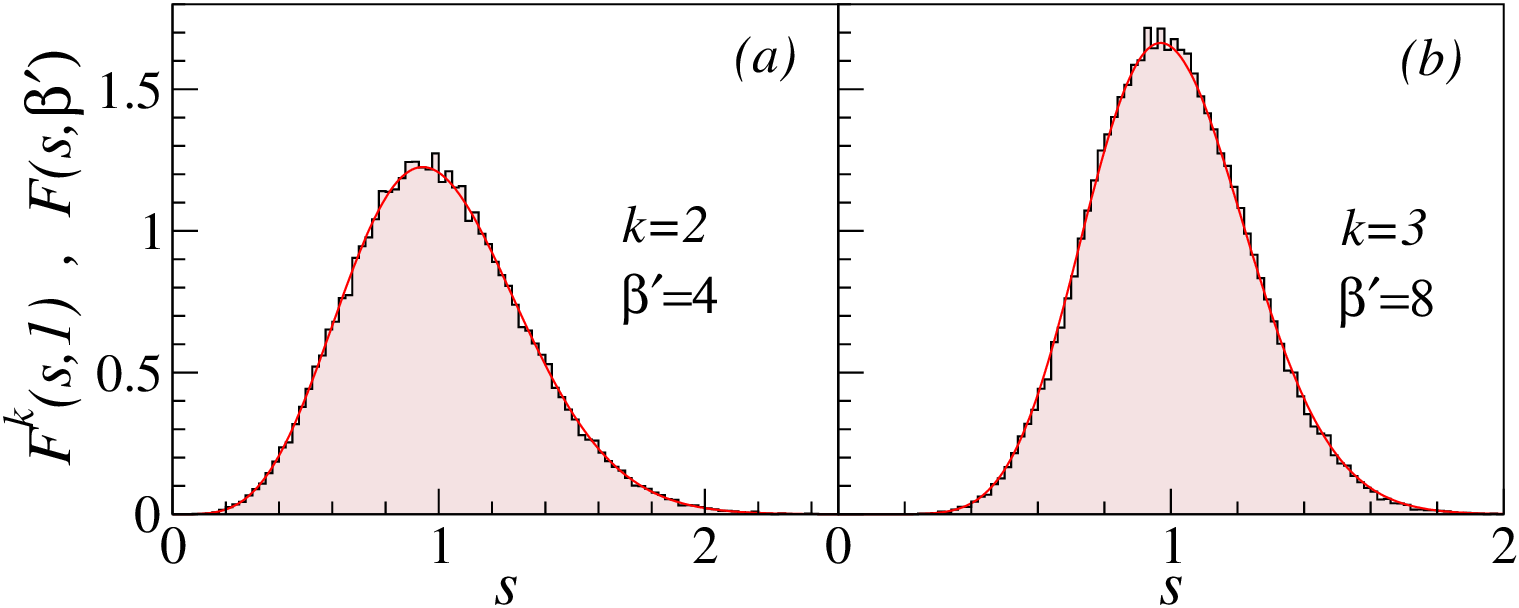}
\caption{The histograms are the $k$-th order spacing distribution 
for the spectra of Wishart matrix. The histograms display a good agreement with $F(s,\beta')$
shown as solid line and $\beta'$ is given by Eq. \ref{eq3}.}
\label{spadist}
\end{figure}

\begin{figure}[b]
\includegraphics*[width=3.4in]{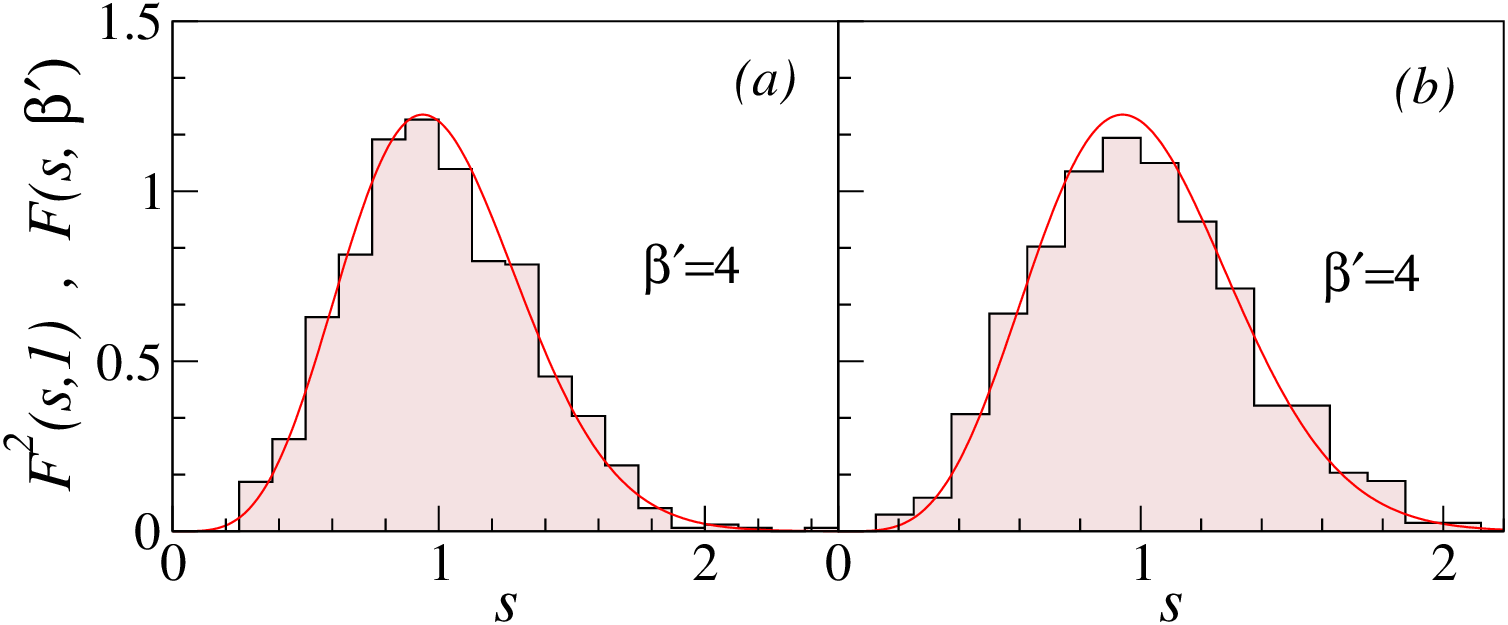}
\caption{The histograms are the $k$-th spacing distribution for the spectra of correlation 
matrix (a) from mean sea level pressure data and (b) from S\&P500 stock market data.
The computed histograms display a good agreement with $F(s,\beta')$ shown as solid 
line and $\beta'$ is given by Eq. \ref{eq3}.}
\label{spadist_sys}
\end{figure}

Further, we consider the time series of monthly mean sea level pressure over the north
Atlantic ocean. The monthly data is drawn from NCEP reanalysis archives \cite{ncep} and is
available over equally spaced latitude/longitude grids for the North Atlantic
region bounded by (0 -- 90$^o$ N, 120$^o$ W -- 30$^o$ E) for the years 1948 to 2017.
Thus, in this case, $N=434$ grid points and $T=840$ months, satisfying the
condition $T/N > 1$. An analysis of the climate phenomenon of north Atlantic
oscillation was performed by constructing an empirical correlation matrix from
this data and using RMT statistics such as the spacing and eigenvector
distributions \cite{Santhanam2001}. In Fig. \ref{atm_fin}(d), the spacing ratio distributions
for the nearest neighbour spacings obtained from the spectra of this correlation
matrix is shown. The computed histogram is seen to be well described by the
theoretical distribution in Eq. \ref{ratio_goe} obtained for Gaussian ensembles.
The higher order spacing ratio distributions shown in Fig. \ref{atm_fin}(e-f)
display a good agreement with $P(r,\beta')$, as anticipated by Eq. \ref{eq3}.

Both these empirical correlation matrix spectra are computed from a relatively
short sequence of time series compared to the length of time series used in
computing Wishart spectra for Fig. \ref{wishartb1}. Hence, the noise level for
the correlations are higher than for the Wishart case, and it is evident
in the higher order spectral statistics shown in Fig. \ref{atm_fin}. This
also manifests as a poor agreement with the $\langle r \rangle$ values shown
in Table I. Finally, we point out that violating the conditions, $N>>1$ and $T \ge N \gg 1$,
leads to deviations from scaling relation Eqs. \ref{eq2}-\ref{eq3} due to
finite size effects as shown in Fig. \ref{fig:k11convergence}. In the top panel of the figure, it is observed that 
the minima of $D(\beta')$ converges to the predicted value $\beta'=76$ on considering a (relatively) small range 
of eigenvalues in the bulk, where the density of states may be assumed to be constant. This finite size or 
convergence effect has also been discussed for Gaussian ensembles in Ref.~\cite{Harshini2018a}.
While in the bottom panel, it is observed that the minima of $D(\beta')$ does not converge to the predicted 
value $\beta'=229$ even on considering a (relatively) small range of eigenvalues in the bulk. To obtain a 
constant local density of states over a larger energy scale, thus requires a random matrix of larger dimensions.


\subsection{Spacing distributions}

We have also studied the validity of the scaling relation in Eq. \ref{eq3} for the more popular eigenvalue spacing 
distribution \cite{rmt1,Forresterbook}. In order to examine this, the $k$-th nearest neighbour spacing is defined as 
$s^{(k)} = (s_{i+k} - s_i)/\langle s \rangle, i=1,2, \dots$. Based on the analytical result 
(see Appendix~\ref{Appendix1}) it is postulated that the second (third) order spacing distribution is 
$F(s,\beta') = A_{\beta'} s^{\beta'} e^{-B_{\beta'} s^2}$, a form that is reminiscent of the Wigner surmise, 
where $\beta'=3\beta+1$ ($\beta'=6\beta+2$). These $\beta'$ agrees with Eq.~\ref{eq3} for $k=2$ ($k=3$).
The constants $A_{\beta'}$ and $B_{\beta'}$ depend on $\beta'$ and are given in Ref.~\cite{GuhrReport}.
In Fig. \ref{spadist}, we verify this claim for the computed Wishart matrix spectra.
The computed histograms display an excellent agreement with the spacing distributions $F(s,\beta'=4)$ 
and $F(s,\beta'=8)$ respectively. Fig. \ref{spadist_sys}(a,b) displays next-nearest neighbour
($k=2$) spacing distribution for the data drawn from mean sea level pressure and S\&P500 stocks.
In both these cases, a good agreement with the anticipated $F(s,\beta')$ is evident.
For $k > 3$, it does not appear straightforward to extend these results due to
pronounced finite size effects and the limitations of pushing the spacing distributions
postulated based on $s \to 0$ results well beyond its regime of validity.

\section{Conclusions}
\label{sec:conclusions}

The empirical correlation matrices are widely used in many areas of sciences
and engineering as tools to extract information from large datasets.
Typically, this is done by analyzing their spectra, the bulk of which are
known to follow the random matrix theory predictions, especially for the
eigenvalue density and the popular nearest neighbour spacing distribution.
Computation of spacing distribution involves unfolding the spectra through an
ambiguous fitting procedure. In recent years, the spacing ratio has 
become a popular alternative to spacing distributions since the former
do not depend on the eigenvalue density and hence it does not require 
unfolding. In this work, for the Wishart matrices of order $N \gg 1$, we focus 
on the higher order spacing statistics and show that 
$k$-th order spacing ratio distribution $P^k(r,\beta)$
can be obtained in terms of the corresponding nearest neighbour $(k=1)$ distribution,
$P(r,\beta')$, with $\beta' > \beta$, and $\beta'$ depends on $k$ and $\beta$.
We have used the correspondence of the Wishart eigenvalues
with the charged particles of a two-dimensional coloumb gas to explain $\beta' > \beta$.
Further, using analytical and simulation results, a similar scaling with a limited
scope is obtained for the spacing distributions of Wishart matrix spectra
and empirical correlation matrices.

We demonstrate the validity of scaling in eigenvalue fluctuations using the spectra 
drawn from an ensemble of Wishart matrices.  As an application with observed datasets,
the scaling in fluctuations is also shown for the spectra of empirical correlation matrices 
obtained from S\&P500 stock market data and mean sea level pressure data over North Atlantic
ocean, both of these had earlier been analysed from RMT point of view.
It would be interesting
to obtain these results exactly for the Wishart ensemble. This opens up new
tests for the claims of universality of eigenvalue fluctuations and further it can
potentially determine the timescales over which RMT-like fluctuations hold good
for empirical correlation matrices.

\section{Acknowledgements} 
Authors thank Dr Giacomo Livan for providing us  S\&P 500 correlation
matrix data for S\&P 500 stocks \cite{livan1}, whose eigenvalues are analyzed in this 
work. UTB acknowledges the funding received from the Department of Science
and Technology, India under the scheme Science and Engineering Research Board (SERB)
National Post Doctoral Fellowship (NPDF) file number PDF/2015/00050.

\appendix
\section{Spacing distributions for second and third order}
\label{Appendix1}

In this Appendix the analytical result leading to the second and third order spacing distributions is derived.
Consider the random Wishart matrix $W$ of order $N$
and specialized to the case of the next-nearest neighbour $(k=2)$ spacing distribution
that can be obtained from Wishart matrix of order $T=3$ with three eigenvalues, 
$\{E_1, E_2, E_3\}$. Then, the jpdf of the eigenvalues 
$E_l \geq 0, l=1,2,3$ for the Wishart ensemble is given as follows:
\begin{equation}
f(\{E_l\})=\frac{1}{W_{a \beta T}}\prod_{i=1}^T E_i^{\beta a/2}e^{-\beta E_i/2}
\prod_{1\leq j<p\leq T} |E_p-E_j|^{\beta},
\nonumber
\end{equation}
where $a=N-T+1-2/\beta$ and $W_{a \beta T}$ is a constant \cite{Forresterbook}. Further, with $T=3$,
$N$ and $\beta$ are choosen such that $a=0$. Then, the jpdf can be obtained as,
\begin{equation}
f(E_1,E_2,E_3)=\frac{3!}{W_{0 \beta 3}}\prod_{i=1}^3 e^{-\beta E_i/2} \prod_{1\leq j<p\leq 3} |E_p-E_j|^{\beta}.
\end{equation}
Using the transformation $x=E_2-E_1$, $y=E_3-E_2$, we obtain $E_2=E_1+x$, $E_3=E_1+x+y$ and
\begin{equation}
f(E_1,x,y)=\frac{3!}{W_{0 \beta 3}}  x^\beta\, y^\beta\, (x+y)^\beta \, e^{-c\beta (3E_1+2x+y)/2}.
\end{equation}
Let $K_1=3!/W_{0 \beta 3}$ and by integrating over $E_1$, one obtains
\begin{equation}
f(x,y)=\frac{2 K_1}{3 \beta c} x^\beta\, y^\beta\, (x+y)^\beta e^{-c\beta (2x+y)/2}.
\end{equation}
It can be seen that $0\leq x+y=E_3-E_1=s$ and $y=s-x$. After some algebra,
the next-nearest-neighbour $(k=2)$ spacing distribution $\widetilde{F}^2(s)$ can be obtained as,
\begin{equation}
\widetilde{F}^2(s) = \frac{s^{3\beta+1} e^{-c\beta s/2}}{2^{-1} K_1^{-1} 3 \beta c} \sum_{q=0}^{\beta} \sum_{n=0}^{\infty}
 {\beta \choose q} \frac{s^n (-1)^{\beta-q} (-c\beta/2)^n}{n! (2\beta-q+n+1)} \\
\end{equation}
In the limit of $s \to 0$, the leading behavior is proportional to $s^{\beta'}$, where $\beta'=3\beta+1$. The 
result derived above can be extended easily for the case of $k=3$ as well, resulting in $\beta'= 6\beta+2$. Thus, 
based on these analytical results and in the spirit of the scaling relation Eqs. \ref{eq2}-\ref{eq3}, it is 
postulated that the second (third) order spacing distribution is 
$F(s,\beta') = A_{\beta'} s^{\beta'} e^{-B_{\beta'} s^2}$
where $\beta'=3\beta+1$ ($\beta'=6\beta+2$). The constants $A_{\beta'}$ and $B_{\beta'}$ 
(given in Ref. \cite{GuhrReport}) depend on $\beta'$ (Eq. \ref{eq3}).

\end{document}